# Probabilistic Models for Unified Collaborative and Content-Based Recommendation in Sparse-Data Environments

**Alexandrin Popescul and Lyle H. Ungar**
Department of Computer and Information Science
University of Pennsylvania
Philadelphia, PA 19104
popescul@unagi.cis.upenn.edu
ungar@cis.upenn.edu

**David M. Pennock and Steve Lawrence**
NEC Research Institute
4 Independence Way
Princeton, NJ 08540
dpennock@research.nj.nec.com
lawrence@research.nj.nec.com

## Abstract

Recommender systems leverage product and community information to target products to consumers. Researchers have developed collaborative recommenders, content-based recommenders, and a few hybrid systems. We propose a unified probabilistic framework for merging collaborative and content-based recommendations. We extend Hofmann's (1999) aspect model to incorporate three-way co-occurrence data among users, items, and item content. The relative influence of collaboration data versus content data is not imposed as an exogenous parameter, but rather emerges naturally from the given data sources. However, global probabilistic models coupled with standard EM learning algorithms tend to drastically overfit in the sparse-data situations typical of recommendation applications. We show that secondary content information can often be used to overcome sparsity. Experiments on data from the ResearchIndex library of Computer Science publications show that appropriate mixture models incorporating secondary data produce significantly better quality recommenders than $k$-nearest neighbors ($k$-NN). Global probabilistic models also allow more general inferences than local methods like $k$-NN.

## 1 INTRODUCTION

The Internet offers tremendous opportunities for mass personalization of commercial transactions. Web businesses ideally strive for global reach, while maintaining the feel of a neighborhood shop where the customers know the owners, and the owners are familiar with the customers and their specific needs. To show a personal face on a massive scale, businesses must turn to automated techniques like so-called *recommender systems* (Resnick & Varian, 1997). These systems suggest products of interest to consumers based on their explicit and implicit preferences, the preferences of other consumers, and consumer and product attributes. For example, a movie recommender might combine explicit ratings data (e.g., Bob rates *X-men* a 7 out of 10), implicit data (e.g., Mary purchased *Hannibal*), user demographic information (e.g., Mary is female), and movie content information (e.g., *Mystery Men* is a comedy) to make recommendations to specific users.

Traditionally, recommender systems have fallen into two main categories. *Collaborative filtering* methods utilize explicit or implicit ratings from many users to recommend items to a given user (Breese et al., 1998; Resnick et al., 1994; Shardanand & Maes, 1995). *Content-based* or *information filtering* methods make recommendations by matching a user's query, or other user information, to descriptive product information (Mooney & Roy, 2000; Salton & McGill, 1983). Pure collaborative systems tend to fail when little is known about a user, or when he or she has uncommon interests. On the other hand, content-based systems cannot account for community endorsements; for example, an information filter might recommend *The Mexican* to a user who likes Brad Pitt and Julia Roberts, even though many like-minded users strongly dislike the film. Several researchers are exploring hybrid collaborative and content-based recommenders to smooth out the disadvantages of each (Basu et al., 1998; Claypool et al., 1999; Good et al., 1999).

In this paper, we propose a generative probabilistic model for combining collaborative and content-based recommendations in a normative manner. The model builds on previous two-way co-occurrence models for information filtering (Hofmann, 1999) and collaborative filtering (Hofmann & Puzicha, 1999). Our model incorporates three-way co-occurrence data by presuming that users are interested in a set of latent topics which in turn "generate" both items and item content information. Model parameters are learned using *expectation maximization* (EM), so the relative contributions of collaborative and content-based data are determined in a sound statistical manner. When data is ex-tremely sparse, as is typically the case for collaboration data, EM can suffer from overfitting. In Sections 4 and 5, we present two techniques to effectively increase the density of the data by exploiting secondary data. The first uses a similarity measure to fill in the user-item co-occurrence matrix by inferring which items users are likely to have accessed without the system's knowledge. The second creates an implicit user-content co-occurrence matrix by treating each user's access to an item as if it were many accesses to all of the pieces of content in the item's descriptive information. We evaluate these models in the context of a document recommendation system. Specifically, we train and test the models on data from ResearchIndex,[1] an online digital library of Computer Science papers (Lawrence et al., 1999; Bollacker et al., 2000). Section 6 presents empirical results and evaluations. In Section 6.2, we demonstrate the potential ineffectiveness of EM in sparse-data situations, using both ResearchIndex data and synthetic data. In Section 6.3, we show that both of our density-augmenting methods are effective at reducing overfitting and improving predictive accuracy. Our models yield more accurate recommendations than the commonly-employed $k$-nearest neighbors ($k$-NN) algorithm. Moreover, our global models can produce predictions for any user-item pair, whereas local methods like $k$-NN are simply incapable of producing meaningful recommendations for many user-item combinations.

## 2 BACKGROUND AND RELATED WORK

A variety of collaborative filtering algorithms have been designed and deployed. The Tapestry system relied on each user to identify like-minded users manually (Goldberg et al., 1992). GroupLens (Resnick et al., 1994) and Ringo (Shardanand & Maes, 1995), developed independently, were the first to automate prediction. Typical algorithms compute similarity scores between all pairs of users; predictions for a given user are generated by weighting other users' ratings proportionally to their similarity to the given user. A variety of similarity metrics are possible, including correlation (Resnick et al., 1994), mean-squared difference (Shardanand & Maes, 1995), vector similarity (Breese et al., 1998), or probability that users are of the same type (Pennock et al., 2000b). Other algorithms construct a model of underlying user preferences, from which predictions are inferred. Examples include Bayesian network models (Breese et al., 1998), dependency network models (Heckerman et al., 2000), clustering models (Ungar & Foster, 1998), and models of how people rate items (Pennock et al., 2000b). Collaborative filtering has also been cast as a machine learning problem (Basu et al., 1998; Billsus & Pazzani, 1998; Nakamura & Abe, 1998) and as a list-ranking problem (Cohen et al., 1999; Freund et al., 1998; Pennock et al., 2000a). Singular Value Decomposition (SVD) was used to improve scalability of collaborative filtering systems by dimensionality reduction (Sarwar et al., 2000).

Pure information filtering systems use only content to make recommendations. For example, search engines recommend web pages with content similar to (e.g., containing) user queries (Salton & McGill, 1983). In contrast to collaborative methods, content-based systems can even recommend new (previously unaccessed) items to users without any history in the system. Mooney & Roy (2000) develop a content-based book recommender using information extraction and machine learning techniques for text categorization.

Several authors suggest methods for combining collaborative filtering with information filtering. Basu et al. (1998) present a hybrid collaborative and content-based movie recommender. Collaborative features (e.g., Bob and Mary like *Titanic*) are encoded as set-valued attributes. These features are combined with more typical content features (e.g., *Traffic* is rated R) to inductively learn a binary classifier that separates liked and disliked movies. Also in a movie recommender domain, Good et al. (1999) suggest using content based software agents to automatically generate ratings to reduce data sparsity. Claypool et al. (1999) employ separate collaborative and content-based recommenders in an online newspaper domain, combining the two predictions using an adaptive weighted average: as the number of users accessing an item increases, the weight of the collaborative component tends to increase. Web hyperlinks and document citations can be thought of as implicit endorsements or ratings. Cohn and Hofmann (2001) combine document content information with this type of connectivity information to identify principle topics and authoritative documents in a collection.

Recommender systems technology is in current use in many Internet commerce applications. For example, the University of Minnesota's GroupLens and MovieLens[2] research projects spawned Net Perceptions,[3] a successful Internet startup offering personalization and recommendation services. Alexa[4] is a web browser plug-in that recommends related links based in part on other people's web surfing habits. A growing number of companies,[5] including Amazon.com, CDNow.com, and Levis.com, employ or provide recommender system solutions (Schafer et al., 1999). Recommendation tools originally developed at Microsoft Research are now included with the Commerce Edition of Microsoft's SiteServer,[6] and are currently in use at multiple

---

[1] http://researchindex.org/
[2] http://movielens.umn.edu/
[3] http://www.netperceptions.com/
[4] http://www.alexa.com/
[5] http://www.cis.upenn.edu/~ungar/CF/
[6] http://www.microsoft.com/siteserver

sites.

## 3  THREE-WAY ASPECT MODEL

Hofmann (1999) proposes an *aspect model*—a latent class statistical mixture model—for associating word-document co-occurrence data with a set of latent variables. Hofmann and Puzicha (1999) apply the aspect model to user-item co-occurrence data for collaborative filtering. In the context of a document recommender system, users $u \in U = \{u_1, u_2, \ldots, u_N\}$, together with the documents they access $d \in D = \{d_1, \ldots, d_M\}$, form observations $(u, d)$, which are associated with one of the latent variables $z \in Z = \{z_1, \ldots, z_K\}$. Conceptually, the latent variables are topics. Users choose among topics according to their interests; topic variables in turn "generate" documents. Users are assumed independent of documents, given the topics. The joint probability distribution over users, topics, and documents is $\Pr(u) \Pr(z|u) \Pr(d|z)$. An equivalent specification of the joint distribution that treats users and documents symmetrically is $\Pr(z) \Pr(u|z) \Pr(d|z)$. The joint distribution over just users and documents is

$$\Pr(u, d) = \sum_z \Pr(z) \Pr(u|z) \Pr(d|z).$$

Model parameters are learned using EM (or variants) to find a local maximum of the log-likelihood of the training data. After the model is learned, documents can be ranked for a given user according to $\Pr(d|u) \propto \Pr(u, d)$; that is, according to how likely it is that the user will access the corresponding document. Documents with high $\Pr(d|u)$ that the user has not yet seen are good candidates for recommendation. Note that the aspect model allows multiple topics per user, unlike most clustering algorithms that assign each user to a single class.

This model is a pure collaborative filtering model; document content is not taken into account. We propose an extension of the aspect model to include three-way co-occurrence data among users, documents, and document content. An observation is a triple $(u, d, w)$ corresponding to an event of a user $u$ accessing document $d$ containing word $w$. Conceptually, users choose (latent) topics $z$, which in turn generate both documents and their content words. Users, documents, and words are assumed independent, given the topics. An asymmetric specification of the joint distribution corresponding to this conceptual viewpoint is $\Pr(u) \Pr(z|u) \Pr(d|z) \Pr(w|z)$. Figure 1 depicts this model as a Bayesian network. An equivalent symmetric specification (obtained by reversing the arc from users to topics) is $\Pr(z) \Pr(u|z) \Pr(d|z) \Pr(w|z)$. Marginalizing out $z$, we obtain

$$\Pr(u, d, w) = \sum_z \Pr(z) \Pr(u|z) \Pr(d|z) \Pr(w|z).$$

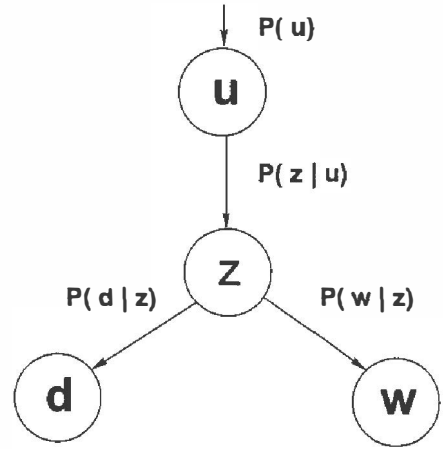

Figure 1: Graphical representation of the three-way aspect model.

Let $n(u, d, w)$ be the number of times user $u$ "saw" word $w$ in document $d$. That is, $n(u, d, w) = n(u, d) \times n(d, w)$, where $n(u, d)$ is the number of times user $u$ accessed document $d$, and $n(d, w)$ is the number of times word $w$ occurs in document $d$. Given training data of this form, the log likelihood $L$ of the data is

$$L = \sum_{u,d,w} n(u, d, w) \log \Pr(u, d, w).$$

The corresponding EM algorithm is:

E step:

$$\Pr(z|u, d, w) = \frac{\Pr(z) \Pr(u|z) \Pr(d|z) \Pr(w|z)}{\sum_{z'} \Pr(z') \Pr(u|z') \Pr(d|z') \Pr(w|z')}$$

M step:

$$\Pr(u|z) \propto \sum_{d,w} n(u, d, w) \Pr(z|u, d, w)$$

$$\Pr(d|z) \propto \sum_{u,w} n(u, d, w) \Pr(z|u, d, w)$$

$$\Pr(w|z) \propto \sum_{u,d} n(u, d, w) \Pr(z|u, d, w)$$

$$\Pr(z) \propto \sum_{u,d,w} n(u, d, w) \Pr(z|u, d, w)$$

The E and M steps are repeated alternately until a local maximum of the log-likelihood is reached.

As in the two-way model, $\Pr(d|u) \propto \sum_w \Pr(u, d, w)$ is used to recommend documents to users. Both content and collaboration data can influence recommendations. The relative weight of each type of data depends on the nature of the given data; EM automatically exploits whatever data source is most informative.

Hofmann (1999) proposes a variant of EM called *tempered EM* (TEM) to help avoid overfitting and improve general-ization. TEM makes use of an inverse computational temperature $\beta$. EM is modified by raising the conditionals in the right-hand side of the E step equation to the power $\beta$. TEM starts with $\beta = 1$, and decreases $\beta$ with the rate $\eta < 1$ using $\beta = \beta \times \eta$, when the performance on a held-out portion of the training set deteriorates.

In Section 6.2, we see that even TEM fails to generalize when data is extremely sparse. In the next two sections, we propose two methods that effectively increase data density, thereby improving learning performance.

## 4 SIMILARITY-BASED DATA SMOOTHING

One approach to overcoming the overfitting problem with sparse data is to use the similarity between items to smooth the co-occurrence data matrix. The co-occurrence matrix contains integer entries that are the number of times the corresponding row and column items co-occur in the observed data set. Similarity between items in the database can be used to fill some zeros in the co-occurrence data matrix, thus reducing sparsity and helping to address overfitting. Consider a user $u$ who has accessed document $d_i$ once, and assume there exists a document $d_j$ that has not been accessed by $u$, and that documents $d_i$ and $d_j$ are very similar in content (e.g., they share many words in common). Consider a similarity metric which yields $sim(d_i, d_j) = 0.7$. Informally, we may believe that there is a 70% chance that user $u$ actually has seen document $d_j$, even though the system does not know it. Using this reasoning, we propose to preprocess the initial co-occurrence data matrix, by filling in some of the zeros with the aggregate similarity between the corresponding document and the documents definitely seen by user $u$. The co-occurrence matrix will no longer be integer valued, but may also contain similarity values which range between 0 and 1. The EM algorithm used in the original aspect model also converges in this situation.

The most frequently used similarity measure in information retrieval is vector-space cosine similarity (Salton & McGill, 1983). Each document is viewed as a vector whose dimensions correspond to words in the vocabulary; the component magnitudes are the *tf-idf* weights of the words. *Tf-idf* is the product of term frequency $tf(w, d)$—the number of times word $w$ occurs in the corresponding document $d$—and inverse document frequency

$$idf(w) = \log \frac{|D|}{df(w)},$$

where $|D|$ is the number of documents in a collection and $df(w)$ is the number of documents in which word $w$ occurs at least once. The similarity between two documents is then

$$sim(\mathbf{d_i}, \mathbf{d_j}) = \frac{\mathbf{d_i} \cdot \mathbf{d_j}}{||\mathbf{d_i}|| ||\mathbf{d_j}||},$$

where $\mathbf{d_i}$ and $\mathbf{d_j}$ are vectors with *tf-idf* coordinates as described above.

In our setting, the user-document co-occurrence data matrix is smoothed by replacing zero entries with average similarities above a certain threshold between the corresponding document and all documents that the user has accessed. This effectively increases the density (i.e., the fraction of non-zero entries) in the matrix. Figure 2 shows how the density of the ResearchIndex data (described in detail in Section 6.1) changes depending on the similarity threshold used in smoothing.

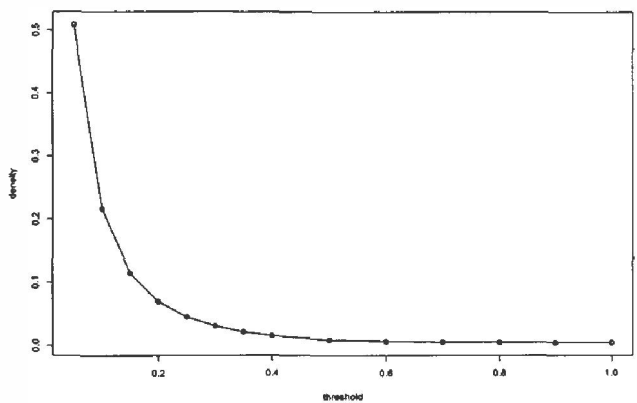

Figure 2: Density of the data against the similarity threshold used in smoothing.

## 5 IMPLICIT USER-WORDS ASPECT MODEL

As another method to overcome overfitting due to sparsity, we propose a model where the co-occurrence data points represent events corresponding to users looking at words in a particular document. The concept of a document is removed to create observations $(u, w)$. Sparsity is drastically reduced because documents contain many words, and many words are contained in multiple documents.

In this case, the aspect model produces estimates of conditional probabilities $\Pr(u|z)$ and $\Pr(w|z)$, as well as the latent class variable priors $\Pr(z)$, allowing us to compute

$$\Pr(u, w) = \sum_z \Pr(z) \Pr(u|z) \Pr(w|z).$$

But we are still interested in estimating probabilities $\Pr(d|u)$ to produce recommendations of the papers that have the highest scores on the $\Pr(d|u)$ scale for a given user $u$. By assuming conditional independence of words in a document, we can overcome this problem by treating a document as a bag of words: the probability of a document is the product of the probabilities of the words it contains, adjusted for different document lengths with the geometric

mean:

$$\Pr(d|u) \propto \left(\prod_i \Pr(w_i|u)\right)^{1/|d|},$$

where $w_i$ are words in $d$ and $|d|$ is the length of $d$. Conditional probabilities $\Pr(w_i|u)$ follow directly from the model:

$$\Pr(w_i|u) = \frac{\Pr(u, w_i)}{\sum_w \Pr(u, w)}.$$

Inclusion of words through documents, and eliminating documents from direct participation in modeling, increased the density of our dataset (described below) from 0.38% to almost 9%.

# 6 RESULTS AND EVALUATION

Section 6.1 describes the ResearchIndex data. In Section 6.2, we examine under what conditions learning occurs at all, by measuring the increase in the log-likelihood of test data as EM proceeds. We find that if data is too sparse, neither EM nor TEM succeeds in significantly increasing the test data log-likelihood over a random initial guess. In Section 6.3, we evaluate the recommendations of our density-augmented models, according to Breese et al.'s (1998) *rank scoring* metric.

## 6.1 RESEARCHINDEX DATA

The data for our experiments was taken from ResearchIndex (formerly CiteSeer), the largest freely available database of scientific literature (Lawrence et al., 1999; Bollacker et al., 2000). ResearchIndex catalogs scientific publications available on the web in PostScript and PDF formats. The full-text of documents as well as the citations made in them are indexed. ResearchIndex supports keyword-based retrieval and browsing of the database, for example by following the links between papers formed by citations. Document detail page access information was obtained for July to November, 2000 (multiple accesses by the same user were included). Heuristics were used to filter out robots. Words from the first 5 kbytes of the text of each document were extracted.

We used the data from July to October as the training set, and the data from November as the testing set. Due to the rapid growth in usage of ResearchIndex, November accounted for 31% of the total five month activity. The data included 33,050 unique users accessing the details of 177,232 documents. Density of this dataset was only 0.01%.

We extracted a relatively dense (0.38%) subset of the 1000 most active users and the 5000 documents they accessed the most. We believe these very low density levels are typical of many real-world recommendation applications. Experiments reported in this paper were conducted using the relatively dense subset of 1,000 users and 5,000 papers.

## 6.2 OVERFITTING

### 6.2.1 User-Document And User-Document-Word Aspect Models

Training the two-way user-document aspect model on the relatively dense set of 1000 users and 5000 documents resulted in immediate overfitting of EM, meaning that the test data log-likelihood began to fall after only the first or second iteration. This immediate overfitting occured for numbers of latent classes ranging from 3 to 50. Using tempered EM (under several reasonable temperature change schedules) only kept the test data log-likelihood approximately at the same level as the initial random seed, without significant improvements.

Including the words contained in the 5,000 documents, and fitting the three-way aspect model also resulted in immediate overfitting. Again, TEM failed to yield significant improvements in the test data log-likelihood.

### 6.2.2 Standard Aspect Model, Synthetic Data

To examine whether this extreme overfitting was specific to the ResearchIndex data, we tested the aspect model on a simple synthetic data set. Users are divided into three disjoint groups according to the following scheme:

1. users 0–49 read papers 0–299,

2. users 50–99 read papers 300–599, and

3. users 100–149 read papers 600–899,

where the probabilities that users read papers in their interest set are uniform.

We designed the data so that the "correct" model with three latent states is obvious. We generated several datasets of differing densities and trained a three-latent-variable aspect model on each to see whether EM converges to the correct model. We performed validation tests at each iteration with test sets of the same density as the corresponding training set. Figure 3 plots the iteration (averaged over fifty random restarts of EM) where overfitting[7] first occurs versus the dataset density. In datasets of density less than 1.5% the process consistently overfits from the first iteration. For datasets of density 2.5%, test performance begins to deteriorate after about five iterations on average. For datasets of density 4%, overfitting begins after ten iterations.

---

[7]Defined as the point where test data log-likelihood starts deteriorating.

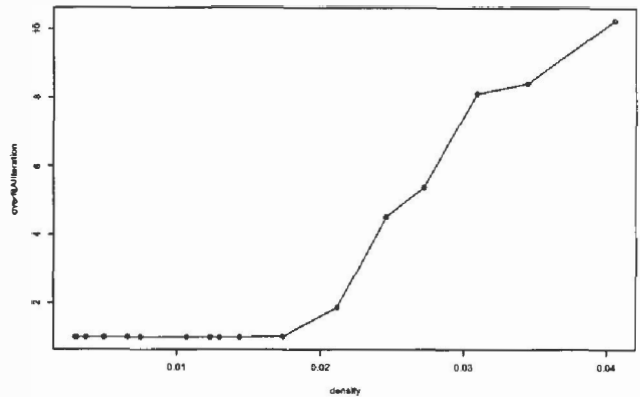

Figure 3: Iteration (averaged over fifty random restarts) where overfitting occurs versus density of the synthetic data.

### 6.3 RECOMMENDATION ACCURACY

We find that both EM and TEM fail on very sparse data, including ResearchIndex data and synthetic data. In contrast, EM *is* effective on both of our density-augmented models (Sections 4 and 5). Here we compare these two models to the $k$-NN algorithm, commonly employed in commercial recommender systems. We use the *rank scoring* metric (Breese et al., 1998) to evaluate recommendations.

#### 6.3.1 Evaluation Criteria

Breese et al. (1998) define the expected utility of a ranked list of items as

$$R_u = \sum_j \frac{\delta(u,j)}{2^{(j-1)/(\alpha-1)}},$$

where $j$ is the rank of an item in the full list of suggestions proposed by a recommender, $\delta(u,j)$ is 1 if user $u$ accessed item $j$ in the test set and 0 otherwise, and $\alpha$ is the viewing half-life, which is the place of an item in the list such that it has a 50% chance of being viewed.[8] As in their paper, we use $\alpha = 5$, and found that our resulting conclusions were not sensitive to the precise value of this parameter. The final score reflecting the utilities of all users in the test set is

$$R = 100 \frac{\sum_u R_u}{\sum_u R_u^{max}},$$

where $R_u^{max}$ is the maximum possible utility obtained when all items that user $u$ has accessed appear at the top of the ranked list.

#### 6.3.2 $k$-Nearest Neighbors

Figure 4 gives $R$ scores for the experiments with $k$-NN in standard formulation on the user-document data for different values of $k$, ranging from 10 to 60 with an interval of 5.

[8]We modify Breese et al.'s formula slightly for the case of observed accesses rather than ratings.

The maximum $R$ value achieved in these experiments was 1.87 for $k = 25$. $R$ scores have local maxima, suggesting their sensitivity to the sparsity of the user-document data.

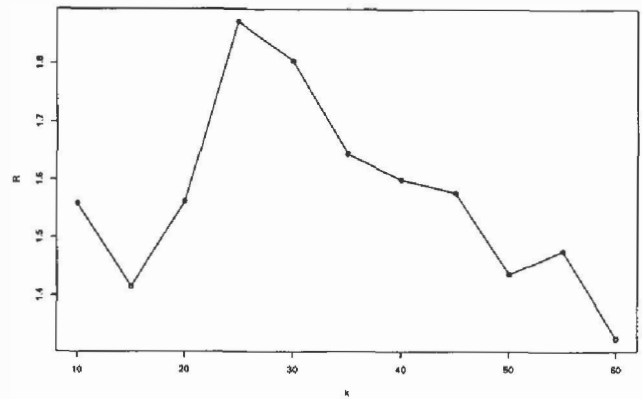

Figure 4: Total utility of the ranked lists over all users produced by $k$-NN.

#### 6.3.3 Smoothed Aspect Model

Figure 5 shows the total utility of the ranked lists ($R$) for all users against the similarity threshold used for smoothing for the example of 25 latent variables. Although the values of $R$ fluctuate, the pattern is clear through the significant linear least squares fit ($p$-value of the slope coefficient is 0.02)—$R$ is larger when more content is included (smaller similarity threshold). As the similarity threshold grows, the initial data matrix becomes sparser, until it becomes impossible to learn (immediate overfitting). Local fluctuations are due to the stochastic nature of EM; in particular, its sensitivity to the randomly initialized parameter values and the number of restarts attempted (five in these experiments) when the data matrix becomes sparser as the similarity threshold grows.

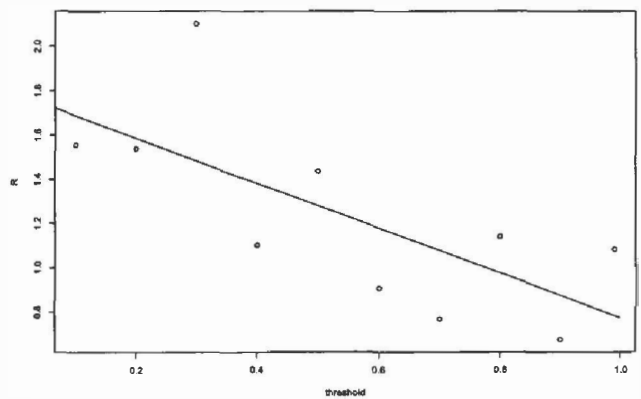

Figure 5: Total utility of the ranked lists over all users produced by the similarity-based User-Document model against the similarity threshold used in smoothing (25 latent class variables).

The maximum value $R$ has reached is 2.10, which is greater than the best $k$-NN result (1.87), but not as good as the User-Words model (2.92), discussed below.

#### 6.3.4 User-Words Aspect Model

Figure 6 shows the $R$ scores for the User-Words aspect model recommender. Experiments include models with the number of hidden class variables $z$ ranging from 10 to 60 with an interval of 10 (two restarts were performed for each experiment). The maximum $R$ value achieved in these experiments is 2.92 for the model with 50 hidden class variables, which is significantly higher than 1.87, the best $R$ value achieved with $k$-NN algorithm.

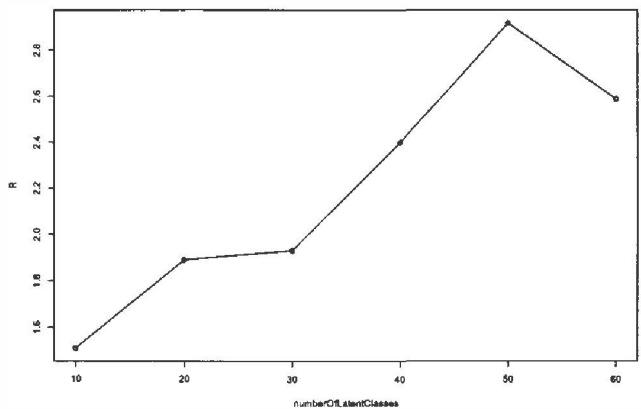

Figure 6: Total utility of the ranked lists over all users produced by the User-Words aspect model.

## 7 CONCLUSIONS AND FUTURE WORK

We presented three probabilistic mixture models for recommending items based on collaborative and content-based evidence merged in a unified manner. Incorporating content into a collaborative filtering system can increase the flexibility and quality of the recommender. Moreover, when data is extremely sparse—as is typical in many real-world applications—additional content information seems almost necessary to fit global probabilistic models at all. The density of ResearchIndex data is only 0.01%. Even the most active users reading the most popular articles induce a subset of density only 0.38%, still too sparse for the straightforward EM and TEM approaches to work. We find that a particularly good way to include content information in the context of a document recommendation system is to treat users as reading words of the document, rather than the document itself. In our case, this increased the density from 0.38% to almost 9%, resulting in recommendations superior to $k$-NN.

There are many areas for future research. Similar methods to those presented here might be used to recommend items such as movies which have attributes other than text. A movie can be viewed as consisting of the director and the actors in it, just as a document contains words. Both of our sparsity reduction techniques, similarity-based smoothing and an equivalent of a user-words aspect model, can be used.

EM is guaranteed to reach only a local maximum of the training data log-likelihood. Multiple restarts need to be performed if one desires a higher quality model. We are planning to investigate ways to intelligently seed EM to reduce the need for multiple restarts, which can be costly when fitting datasets of non-trivial size.

The user-words model does not explicitly use the popularity of items. Including such information may further improve the quality of the recommendations made by the model, but requires additional work on combining and calibrating model predictions with document popularity.

Finally, predictive accuracy was used to validate our models in this paper. We are planning to deploy our recommenders in ResearchIndex and perform a user study collecting information on which recommendations are actually followed by users.